\documentclass[%
 reprint,
 superscriptaddress,
 amsmath,amssymb,
 aps,
 prl,
]{revtex4-2}

\usepackage{graphicx}
\usepackage{listings}
\usepackage{longtable}
\usepackage{color}
\usepackage[colorlinks,allcolors=blue]{hyperref}
\usepackage[dvipsnames]{xcolor}
\usepackage{ulem}
\usepackage[acronym]{glossaries}

\definecolor{dkgreen}{rgb}{0,0.6,0}
\definecolor{gray}{rgb}{0.5,0.5,0.5}
\definecolor{mauve}{rgb}{0.58,0,0.82}
\makeglossary
\newacronym{jj}{JJ}{Josephson junction}
\newacronym{jj_abs}{JJ}{Josephson junction}
\newacronym{cpw}{CPW}{co-planar waveguide}
\newacronym{jer}{JER}{junction-embedded resonator}
\newacronym{jer_abs}{JER}{junction-embedded resonator}
\newacronym{gama}{$\Gamma$}{intrinsic dissipation rate}
\newacronym{eli}{ELI}{effective linear inductance}
\newacronym{Qi}{$Q_{\mathrm{i}}$}{intrinsic quality factor}
\newacronym{np}{$\langle n_{\mathrm{p}} \rangle$}{average photon number}
\newacronym{ltj}{$L_{\mathrm{TJ}}$}{total ELI of JJs}

\newcommand{\thefont}{\expandafter\string\the\font}

\newcommand{\fth}{$f_{\mathrm{2H}}$}
\newcommand{\foh}{$f_{\mathrm{1H}}$}
\newcommand{\ltj}{$L_{\mathrm{TJ}}$}
\newcommand{\atj}{$A_{\mathrm{TJ}}$}
\newcommand{\gamao}{$\Gamma_{\mathrm{0}}$}
\newcommand{\gamaext}{$\Gamma_{\mathrm{ext}}$}
\newcommand{\gamalp}{$\Gamma_{\mathrm{LP}}$}
\newcommand{\gamatj}{$\Gamma_{\mathrm{TJ}}$}
\newcommand{\vgamaint}{$\sim$$1.61\pm0.08\times10^{-8}$}
\newcommand{\vgamaext}{$\sim$$1.61\pm0.16\times10^{-6}$}

\begin{document}

\title{
    Identify and Quantify Various Dissipation Mechanisms of Josephson Junction
    in Superconducting Circuits
}

\author{Hao Deng}
\email{denghao1@shanghaitech.edu.cn}
\affiliation{School of Physical Science and Technology, ShanghaiTech University, Shanghai 201210, China}
\affiliation{School of Information Science and Technology, ShanghaiTech University, Shanghai 201210, China}

\author{Huijuan Zhan}
\affiliation{Quantum Science Center of Guangdong-Hong Kong-Macao Greater Bay Area, Shenzhen 518045, China}
\author{Lijuan Hu}
\affiliation{Quantum Science Center of Guangdong-Hong Kong-Macao Greater Bay Area, Shenzhen 518045, China}
\author{Hui-Hai Zhao}
\affiliation{Zhongguancun Laboratory, Beijing, China}

\author{Ran Gao}
\affiliation{Quantum Science Center of Guangdong-Hong Kong-Macao Greater Bay Area, Shenzhen 518045, China}
\affiliation{Z-Axis Quantum}
\author{Kannan Lu}
\affiliation{Quantum Science Center of Guangdong-Hong Kong-Macao Greater Bay Area, Shenzhen 518045, China}
\affiliation{Z-Axis Quantum}
\author{Xizheng Ma}
\affiliation{Quantum Science Center of Guangdong-Hong Kong-Macao Greater Bay Area, Shenzhen 518045, China}
\author{Zhijun Song}
\affiliation{Shanghai E-Matterwave Sci \& Tech Co., Ltd., Shanghai 201100, China}
\author{Fei Wang}
\affiliation{Quantum Science Center of Guangdong-Hong Kong-Macao Greater Bay Area, Shenzhen 518045, China}
\affiliation{Z-Axis Quantum}
\author{Tenghui Wang}
\affiliation{Quantum Science Center of Guangdong-Hong Kong-Macao Greater Bay Area, Shenzhen 518045, China}
\author{Feng Wu}
\affiliation{Zhongguancun Laboratory, Beijing, China}
\author{Tian Xia}
\affiliation{Huaxin Jushu Microelectronics Co., Ltd., Hangzhou, China}
\author{Gengyan Zhang}
\affiliation{Zhejiang Laboratory, Hangzhou 311100, China}
\author{Xiaohang Zhang}
\affiliation{The Institute for Brain Research, Advanced Interfaces and Neurotechnologies, 
Shenzhen Medical Academy of Research and Translation, Shenzhen, 518100 China}

\author{Chunqing Deng}
\email{dengchunqing@quantumsc.cn}
\affiliation{Quantum Science Center of Guangdong-Hong Kong-Macao Greater Bay Area, Shenzhen 518045, China}
\affiliation{Z-Axis Quantum}

\begin{abstract}

    Pinpointing the dissipation mechanisms and evaluating their impacts to the performance of \gls{jj_abs}
    are crucial for its application in superconducting circuits.
    In this work, we demonstrate the \gls{jer_abs} as a platform
    which enables us to identify and quantify various dissipation mechanisms of \gls{jj_abs}.
    \gls{jer_abs} is constructed by embedding \gls{jj_abs} in the middle of an open-circuit,
    $1/2 \, \lambda$ transmission-line resonator.
    When the 1st and 2nd harmonics of \gls{jer_abs} are excited,
    \gls{jj_abs} experiences different boundary conditions,
    and is dominated by internal and external dissipations, respectively.
    We systematically study these 2 dissipation mechanisms of \gls{jj_abs}
    by varying the \gls{jj_abs} area and number.
    Our results unveil the completely different behaviors of these 2 dissipation mechanisms,
    and quantitatively characterize their contributions,
    shedding a light on the direction of \gls{jj_abs} optimization in various applications.

\end{abstract}

\maketitle

\gls{jj} is one of the key components in the superconducting circuits of quantum computing and quantum metrology.
The performance of \gls{jj} in terms of dissipation is crucial to its applications in these fields.
There are various efforts to investigate the dissipation mechanisms of \gls{jj}.
Directly characterizing the performance of qubit containing \gls{jj} is commonly adopted.
In this approach, usually the qubit frequency is swept 
and the strongly coupled defects inside \gls{jj} are indicated by significant degradation of qubit coherence time
or anti-crossing in qubit spectrum at certain frequency
\cite{Simmonds.Phys.Rev.Lett.93.077003, Martinis.Phys.Rev.Lett.95.210503, Weeden.ax.2506.00193}.
Alternatively, applying direct stimulation to \gls{jj}, such as strain \cite{Grabovskij.Science338.232234}
or external electric field \cite{Lisenfeld.npjQuantumInformation5.105},
tunes the particular dissipation source, while the response is probed
by characterizing the qubit spectrum and coherence.
Another approach is directly switching the material and fabrication processes of \gls{jj} component 
such as the junction barrier and superconducting electrodes,
which also provides insights of dissipation mechanisms of \gls{jj}
\cite{Kline.SST.22.015004, Oh.PRB.74.100502, Anferov.PRApp.21.024047, Kim.CommMat.2.98, VanDamme.Nature.634.7479}.
However, the previous studies are usually designed for a specific kind of target defect in each work,
and difficult to isolate the contributions from the non-junction components of the circuit.
Moreover, the conclusions are generally qualitative, without quantitative benchmark for 
comparison and further optimization.

\begin{figure}[htbp]
    \includegraphics[width = 1\columnwidth]{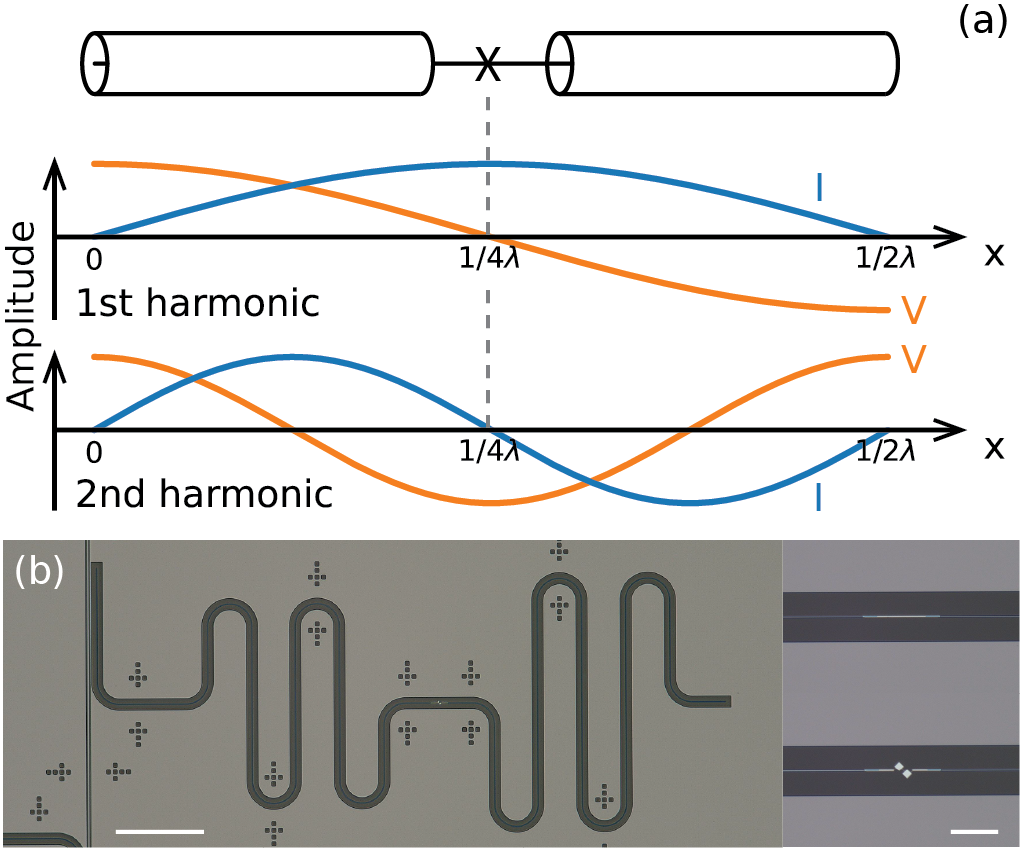}
    \caption{
        (a) The schematic of the \gls{jer}.
        The top panel shows the structure of \gls{jer}.
        The coaxial line represents the transmission-line structure.
        The ``x" mark in the middle is the embedded \gls{jj}.
        The middle and bottom panels demonstrate the voltage (orange curve) and current (blue curve) distributions
        of the 1st and 2nd harmonics along the transmission line of \gls{jer}.
        Note that the embedded \gls{jj} in the middle (i.e., $1/4 \, \lambda$ position)
        experiences different boundary conditions in the 1st and 2nd harmonics.
        (b) The optical microscope image of the \gls{jer},
        where the scale bar indicates 400 $\mathrm{\mu}$m.
        The insets on the right show the zoom-in images
        of the dummy \gls{jj} (top) and 2 \gls{jj}s in series (bottom),
        where the scale bar indicates 50 $\mathrm{\mu}$m.
    }
    \label{fig:diagram}
\end{figure}

In this work, we demonstrate a new method which is able to
identify various dissipation mechanisms of \gls{jj} in-situ,
and quantitatively characterize their contributions.
The approach is through the \gls{jer}:
an open-circuit, $1/2 \, \lambda$, transmission-line resonator
broken by \gls{jj}(s) in the middle [\autoref{fig:diagram}(a)].
In the measurement, we excite the 1st and 2nd harmonics of the resonator, respectively:
in the 1st harmonic, the embedded \gls{jj} locates at the voltage node and current anti-node,
i.e., almost zero potential on the outside of \gls{jj} to the ground plane
while with maximum current flowing inside;
in the 2nd harmonic, the embedded \gls{jj} experiences the voltage anti-node and current node,
exactly opposite to those in the 1st harmonic.
Therefore, we are able to switch completely different boundary conditions to the embedded \gls{jj} in-situ.
In the 1st harmonic, the dissipation of \gls{jj} induced to the resonator
is mainly excited by the current inside, which we refer as the internal dissipation.
Correspondingly, in the 2nd harmonic, the involved dissipation of \gls{jj}
is due to the potential difference (i.e., electric field) between the \gls{jj} and ground plane,
which we refer as the external dissipation.
We systematically characterize the \gls{gama} of a series of \gls{jer}s,
observe that the internal and external dissipations of \gls{jj} have completely different dependences
on the \gls{jj} area and number.
Moreover, we quantitatively extract the net contributions of \gls{jj}'s internal and external dissipations,
without the interference from the non-junction components of the circuit.

In the experiment, we design the \gls{jer} based on the \gls{cpw} resonator.
It should be noticed that, in the \gls{jer} serving for our purpose,
the interruption from the embedded \gls{jj} to the remaining resonator should be as little as possible.
Indeed, in the 1st harmonic, there is a voltage drop across the embedded \gls{jj} depending on its \gls{eli}.
Therefore, to achieve the intended boundary conditions described above,
the \gls{eli} of the embedded \gls{jj} should be
significantly smaller than the effective lumped-element inductance of the remaining resonator.
In the design, on one hand, we choose the geometry of \gls{cpw}
with high characteristic impedance by using a large gap-core ratio,
so that the effective lumped-element inductance of the \gls{cpw} resonator is high enough.
On the other hand, we carefully control the \gls{jj} area and number to make sure that
its contribution does not exceed 15\% of the remaining resonator's effective lumped-element inductance.
Correspondingly, the voltage drop across the \gls{jj} is less than 5\%
of the voltage amplitude at the voltage anti-node of the 1st harmonic.
These efforts guarantee that the embedded \gls{jj} is a perturbation to
voltage/current distribution of the remaining \gls{cpw} resonator.

2 kinds of samples are designed with different variations on \gls{jj}.
In sample A, each \gls{jer} contains a fixed number of 2 \gls{jj}s, but with 3 different areas.
Conversely, in sample B, the \gls{jj} area is fixed in each \gls{jer}, while the number is varied to be 2, 4, and 6.
All the \gls{jj}s in \gls{jer} are connected in series.
Besides, both sample A and sample B contain 1 \gls{jer} with a dummy \gls{jj} 
(an aluminum stripe fabricated at the same time with \gls{jj})
and 2 normal resonators completely made by \gls{cpw} as the controlled devices.
On both samples, the 1st-harmonic frequencies of resonators fall in the range of 5.6-6.1 GHz.
We fabricate our \gls{jer}s with the tantalum film on the sapphire substrate.
Then, the aluminum \gls{jj}s are fabricated with the ``Manhattan'' method
\cite{Potts.IEEProceedingsScienceMeasurementandTechnology148.225228},
and the even number of \gls{jj}s in all \gls{jer}s avoids the parasitic junction at the contact.
All the samples are characterized with the standard power-dependence measurement of resonator
after being cooled down to $\sim$15 mK
\cite{Megrant.AppliedPhysicsLetters100.113510, McRae.ReviewofScientificInstruments91.091101}.
The probing power is carefully clamped to stay away from the nonlinear regime of \gls{jer}s.
See Supplementary Material for more details of the design, fabrication, and measurement setups \cite{Supp.Mate.}.

\begin{figure}[tbp]
    \includegraphics[width = 1\columnwidth]{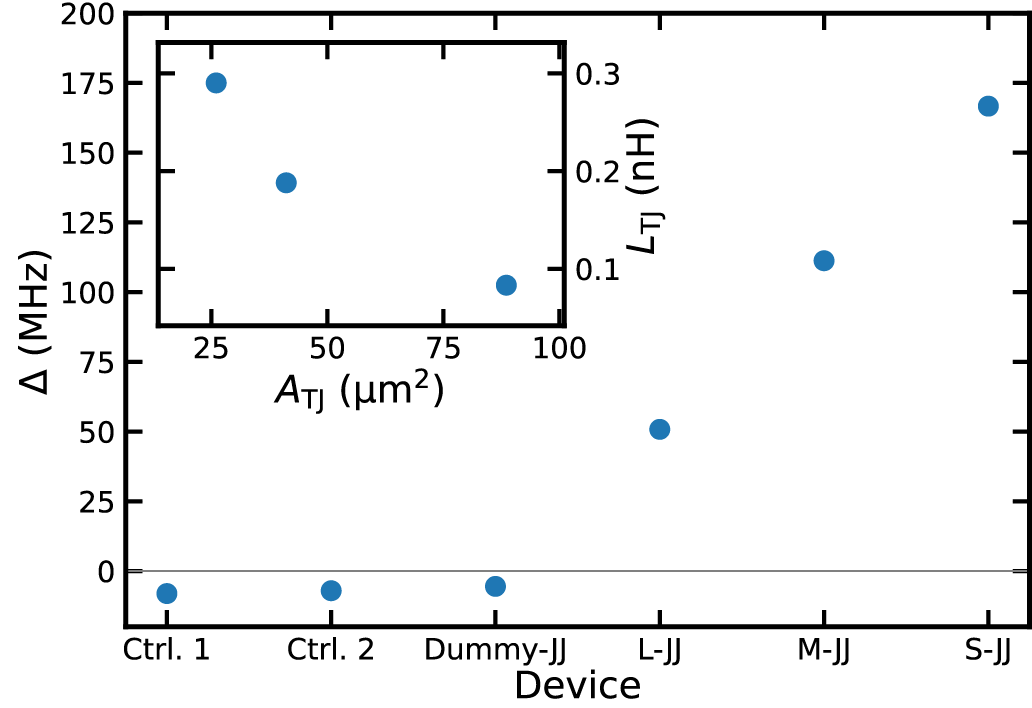}
    \caption{
        $\Delta$ (defined as $1/2 \, f_{\mathrm{2H}} - f_{\mathrm{1H}}$)
        of all the resonators on sample A.
        About the x-axis, Ctrl. 1 and 2 represent 2 controlled devices
        which are normal resonators;
        Dummy-JJ, L-JJ, M-JJ, and S-JJ are the \gls{jer}s containing
        dummy, large-area, medium-area, and small-area \gls{jj}s,
        which are in the ascending order of \gls{eli}.
        The inset shows the extracted total \gls{eli} of \gls{jj}s (\ltj{})
        depending on the total junction area (\atj{}).
    }
    \label{fig:freq}
\end{figure}

We first check the parameters of \gls{jj}s and their involvements in \gls{jer}s
by measuring the frequencies of the 1st and 2nd harmonics.
\autoref{fig:freq} shows the analysis on sample A
through the frequency difference $\Delta = 1/2 \, f_{\mathrm{2H}} - f_{\mathrm{1H}}$,
where $f_{\mathrm{2H/1H}}$ is the 2nd/1st-harmonic frequency.
Ideally, $\Delta$ is zero for a completely isolated, normal resonator.
However, in real case, the resonator couples to the feed line.
\fth{} would become slightly lower than $2 \, f_{\mathrm{1H}}$ 
because of the stronger coupling strength of the 2nd harmonic to the feed line than the 1st harmonic.
Indeed, $\Delta$s of 2 normal resonators (Ctrl. 1 and 2) and the dummy-\gls{jj} \gls{jer}
fall in the slightly negative regime,
and are consistent with the results of circuit model simulation.
In contrast, $\Delta$s of the \gls{jer}s on sample A show clear dependence
on the total \gls{jj} area (\atj{}) which is measured experimentally.
Decreasing \atj{} increases \gls{eli}, resulting in lower \foh{} (but unaffected \fth{}) and larger $\Delta$.
Together with the circuit model simulation by representing \gls{jj} with a linear inductor,
we can extract the \gls{ltj} in each \gls{jer},
which turns out to be close to the design target.
Sample B also shows the same dependence of $\Delta$ on the types of devices,
and reasonable values of the extracted \gls{ltj} (see more details in the Supplementary Material).
Therefore, the frequency data demonstrate that
the embedded \gls{jj}s involve in the \gls{jer}s as our expectation.

\begin{figure}[tbp]
    \includegraphics[width = 1\columnwidth]{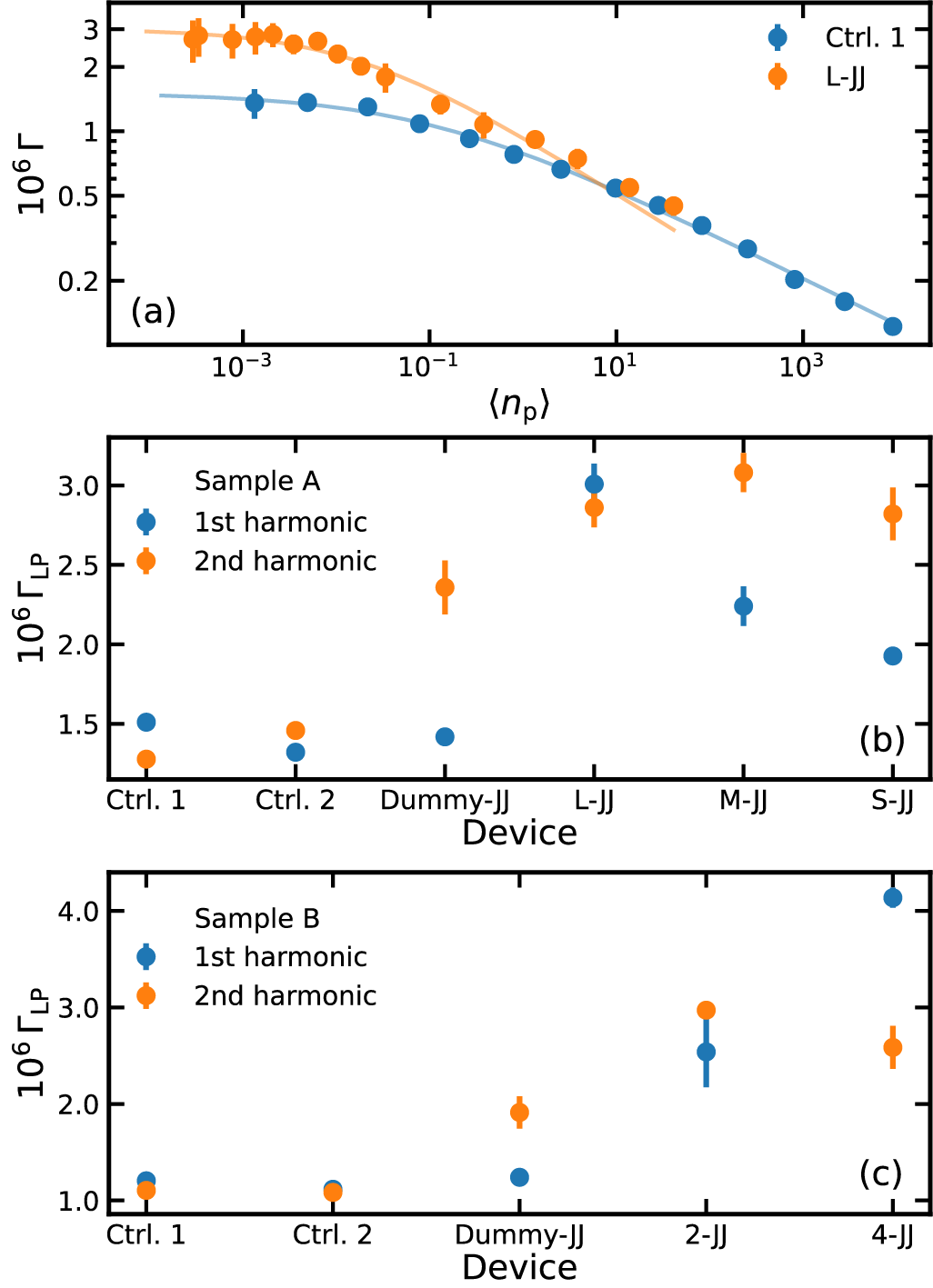}
    \caption{
        (a) Dependences of \gls{gama} on \gls{np} of typical devices on sample A.
        The blue and orange circles represent data points
        of the 1st harmonics of Ctrl. 1 resonator and L-JJ \gls{jer}, respectively.
        The curves in different colors are the fitting results of \autoref{eq:power_dep}
        on the corresponding data points.
        (b) and (c) \gamalp{} of all the devices on sample A and B.
        The blue and orange circles represent the data points of the 1st and 2nd harmonics.
        In (c), we are not able to acquire the data points of the 6-JJ \gls{jer}
        because of its malfunction.
    }
    \label{fig:gama}
\end{figure}

After confirming the involvements of \gls{jj}s in \gls{jer}s,
we characterize the dependence of \gls{gama} (i.e., the reciprocal of intrinsic quality factor)
on the probing power of each resonator.
\autoref{fig:gama}(a) shows the typical results from the devices on sample A.
The probing power is converted to the \gls{np} based on Ref.
\cite{Burnett.JoPCS.969.012131, McRae.ReviewofScientificInstruments91.091101}.
Then, we analyze the data by fitting the dependence of \gls{gama} on \gls{np} with
\cite{Gao.Thesis.2008, McRae.ReviewofScientificInstruments91.091101}:

\begin{equation}
    \Gamma = \Gamma_{\mathrm{0}}\frac{\tanh(hf/2 k_{\mathrm{B}}T)}{\sqrt{1+(\langle n_{\mathrm{p}} \rangle /n_{\mathrm{c}})^{\alpha}}}+\Gamma_{\mathrm{ext}}
    \label{eq:power_dep}
\end{equation}

\noindent where $h$ and $k_{\mathrm{B}}$ are Planck and Boltzmann constants,
$f$ is resonance frequency, $T$ is temperature,
$n_{\mathrm{c}}$, $\alpha$, \gamao{}, and \gamaext{} are fitting parameters.
Particularly, \gamao{} represents the dissipation rate at zero temperature and zero power,
and \gamaext{} indicates the extra dissipation rate independent of probing power.
For the data not reaching saturation at the highest power in our measurement,
\gamaext{} is forced to be 0 for the fitting accuracy.

We focus on the overall, intrinsic dissipation rate of the resonator
represented by the level of saturation in the low-power regime (\gamalp{}).
In the description of \autoref{eq:power_dep}, $\Gamma_{\mathrm{LP}} = \Gamma_{\mathrm{0}} + \Gamma_{\mathrm{ext}}$.
\autoref{fig:gama}(b) demonstrates \gamalp{} values of all the resonators on sample A,
including both the 1st and 2nd harmonics.
We have several observations on the data:
i, 2 controlled devices have almost consistently low \gamalp{}
independent of the orders of harmonics or the individuals,
indicating that the controlled devices provide a reasonably accurate estimation of \gamalp{}
contributed by the non-junction components 
(e.g., tantalum film, sapphire substrate, inter/surfaces, etc.).
ii, The \gls{jer} with dummy \gls{jj} shows different \gamalp{} for different harmonics.
The 1st-harmonic \gamalp{} is consistent with those of the controlled devices,
while the 2nd-harmonic \gamalp{} is significantly higher,
locating at a level between the controlled devices and remaining \gls{jer}s.
iii, the 1st and 2nd harmonics of \gls{jer} with real \gls{jj}s 
have different dependence on junction area.
The 2nd-harmonic \gamalp{} seems to be almost independent of \atj{},
while the 1st-harmonic \gamalp{} obviously has a positive correlation with it.
With regard to the other sample, i.e., sample B,
\gamalp{} data showed in \autoref{fig:gama}(c)
have consistent behaviors with the data of sample A,
except the 1st-harmonic \gamalp{} of \gls{jer}:
\gamalp{} follows the increase of the embedded \gls{jj} number.
But it is obvious that \atj{} also increases with the embedded \gls{jj} number in sample B.
Therefore, in both samples, \gamalp{} of \gls{jer} is consistently correlated with \atj{}.

\begin{figure}[tbp]
    \includegraphics[width = 1\columnwidth]{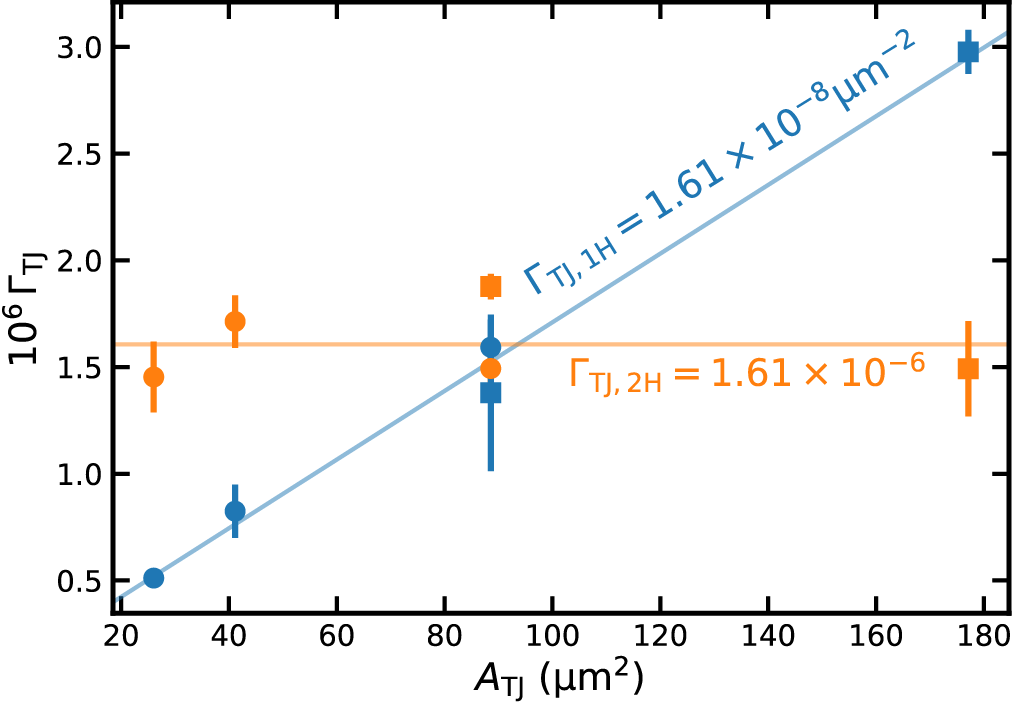}
    \caption{
        Dependence of \gamatj{} on \atj{}.
        The blue and orange colors represent the data points of the 1st and 2nd harmonics.
        The circle and square marks indicate the data points of sample A and B, respectively.
        The blue line is the linear fitting result of the 1st-harmonic \gamatj{} ($\Gamma_{\mathrm{TJ, 1H}}$).
        The orange horizontal line is the averaged value of all the 2nd-harmonic \gamatj{} ($\Gamma_{\mathrm{TJ, 2H}}$).
    }
    \label{fig:gama_atj}
\end{figure}

To verify the role of \atj{} in the dissipation of embedded \gls{jj},
we further analyze the data in the way showed by \autoref{fig:gama_atj}.
To isolate the contributions form the non-junction components,
for both samples and harmonics, we subtract the averaged \gamalp{} of the controlled devices
from the \gamalp{} of each \gls{jer},
and the remaining part is the net contribution of the total embedded \gls{jj} (\gamatj{}).
\autoref{fig:gama_atj} demonstrates the significant difference
between the 1st- and 2nd-harmonic \gamatj{} which is universal to both samples.
The 1st-harmonic \gamatj{} ($\Gamma_{\mathrm{TJ, 1H}}$) shows a clear dependence on \atj{}.
A linear fitting gives $\Gamma_{\mathrm{TJ, 1H}}$ for unit area (1 $\mathrm{\mu m}^{2}$) is \vgamaint{}.
Opposite, the 2nd-harmonic \gamatj{} ($\Gamma_{\mathrm{TJ, 2H}}$) is almost independent of \atj{}.
The averaged $\Gamma_{\mathrm{TJ, 2H}}$ over all of the \gls{jer}s on both samples is \vgamaext{}.
So far, we demonstrate that, 2 different dissipation mechanisms,
i.e., internal and external dissipations
($\Gamma_{\mathrm{TJ, 1H}}$ and $\Gamma_{\mathrm{TJ, 2H}}$ in \autoref{fig:gama_atj}),
identify themselves through completely different dependences on \atj{}.
Moreover, we successfully characterize the contribution of each quantitatively.

We try to discuss the physical origins of the internal and external dissipations of the embedded \gls{jj}
based on our experimental data.
For the external dissipation dominating the 2nd harmonic,
no current flows through \gls{jj},
but there is the strongest electric field between the \gls{jj} and the ground plane nearby.
Therefore, the physical origin of the external dissipation should be the dielectric loss,
which is determined by the electric participation ratio of the dissipative region and its loss tangent
\cite{Wang.APL.107.162601, McRae.ReviewofScientificInstruments91.091101}.
The loss tangent is mainly dominated by the material and fabrication process,
which are identical to all of our devices fabricated on an identical wafer.
The electric participation ratio is mainly tuned by the geometry.
In our design of \gls{jer}, the gap of \gls{cpw} is much larger than
the dimensions of \gls{jj} [see \autoref{fig:diagram}(b) inset].
We expect that the variance of \gls{jj} geometry in our \gls{jer}
has only minor effect to the electric participation ratio.
Therefore, it is reasonable that,
the dielectric loss is almost consistent for the 2nd harmonics of all the \gls{jer}s
regardless the \gls{jj} area and number,
which matches the behavior of external dissipation observed in the experiment.

In the 1st harmonic, \gls{jj} feels the maximum current inside but the lowest electric field outside.
The internal dissipation should be triggered by the current,
which suggests a conductive or inductive property.
There could be several candidates of the internal loss,
such as contact loss, inductive loss of \gls{eli}, etc..
We first rule out the contact loss,
because that such mechanism should not have strong dependence on \atj{}.
Moreover, all the \gls{jer}s have the identical geometry of contact,
including the dummy-\gls{jj} \gls{jer}s of both samples.
But dummy-\gls{jj} \gls{jer}s show the 1st-harmonic \gamalp{} values
almost identical to those of the controlled devices
[see \autoref{fig:gama}(b) and (c)],
indicating that the contact loss is negligible.

If the \gls{eli} is not ideal,
its inductive loss (i.e., as the counterpart of dielectric loss, which could be represented by a magnetic loss tangent) 
could contribute to the internal dissipation of \gls{jj}.
In this case, similar to the dielectric loss, it is expected that the internal dissipation
should have positive correlation with the inductive participation ratio of \gls{jj},
which is defined as the proportion of the \gls{jj} inductive energy among the full energy of \gls{jer}.
The inductive participation ratio roughly scales with \gls{ltj},
i.e., the \gls{jer} containing \gls{jj}s with larger \gls{ltj} suffers from stronger inductive loss of \gls{eli}.
However, this scenario is contradictory to the data of sample A.
\autoref{fig:gama}(b) shows that, the 1st-harmonic \gamalp{} decreases when \atj{} shrinks, i.e., \gls{ltj} increases.
Therefore, the inductive loss of \gls{eli} should not be the dominant source of the internal dissipation
(see more quantitative analysis in Supplementary Material).

Phenomenologically, the internal dissipation is proportional to \atj{}.
Such dependence could suggest a kind of countable defects in \gls{jj}
with an uniform density for all the \gls{jer}s,
and the total number of defects determines the internal dissipation \cite{Bilmes.NPJQI.8.24, ColaoZanuz.PRApp.23.044054}.
This scenario implies that, the \gls{jj} barrier and/or surface could be the host of the defects,
and the two-level systems and/or dangling spins could perform as the countable defects
\cite{Abdurakhimov.PRXQ.3.040332, Braumueller.PRA.13.054079}.
In-depth studies are desired to further unveil the microscopic origin of the internal dissipation.

Besides the identification and quantification of multiple dissipation mechanisms of \gls{jj}, 
our work also sheds a light on the optimization of \gls{jj} in various scenes of applications.
The different behaviors of internal and external dissipations of \gls{jj} 
unveiled by our work provide an important hint:
there could be a transition of dominant dissipation source in \gls{jj} depending on the geometry.
For the device with small \gls{jj} such as transmon, 
the external dissipation is more crucial and the interface between the \gls{jj} electrodes and substrate/vacuum
should be considered seriously \cite{Deng.PRA.19.024013}.
On the contrary, for the device with massive \gls{jj}s such as fluxonium,
the optimization on the \gls{jj}s to reduce the internal dissipation
would bring the most significant gain \cite{Wang.PRApp.23.044064}.
The \gls{jer} demonstrated in our work is an effective tool 
indicating the proper direction of device optimization with quantitative evaluation.

In conclusion, we demonstrate the \gls{jer} as an advantageous platform
to study various dissipation mechanisms of \gls{jj}.
By exciting different harmonics, we can apply different boundary conditions
to the embedded \gls{jj} in-situ,
which stimulate different dominant dissipation mechanisms.
Moreover, by setting the controlled devices,
the contribution from the non-junction components could be isolated.
Through this way, we identify 2 dissipation mechanisms of \gls{jj}.
The external dissipation is independent of \atj{} with an averaged level of \vgamaext{}.
As a comparison, the internal dissipation
shows a significant dependence on \atj{} 
with a dissipation rate of \vgamaint{} $\mathrm{\mu m}^{-2}$,
distinguishing itself from the external one qualitatively.
The external dissipation is consistent with the dielectric loss around \gls{jj},
while the area-scaling behavior of the internal dissipation
suggests an origin of countable defects with an uniform density inside.
Not only serve as a tool to study dissipation mechanisms of \gls{jj},
the \gls{jer} could also perform as a benchmark of \gls{jj} performance,
pointing out the direction of device optimization in various applications.

\section{Acknowledgement}
We thank the former DAMO Quantum Laboratory, Alibaba Group for the support on experiments.
Huijuan Zhan, Lijuan Hu, Ran Gao, Kannan Lu, Xizheng Ma, Fei Wang, Tenghui Wang, and Chunqing Deng
acknowledge the support from Guangdong Provincial Quantum Science Strategic Initiative 
(Grant No. GDZX2407001).
Xiaohang Zhang acknowledges the support from National Natural Science Foundation of China 
(Grant No. 12303096).
Hui-Hai Zhao and Feng Wu are supported by Zhongguancun Laboratory.
Special thanks to Huijuan Zhan and Lijuan Hu, 
who completed the fabrication in an extremely difficult time.


\begin{thebibliography}{23}%
\makeatletter
\providecommand \@ifxundefined [1]{%
 \@ifx{#1\undefined}
}%
\providecommand \@ifnum [1]{%
 \ifnum #1\expandafter \@firstoftwo
 \else \expandafter \@secondoftwo
 \fi
}%
\providecommand \@ifx [1]{%
 \ifx #1\expandafter \@firstoftwo
 \else \expandafter \@secondoftwo
 \fi
}%
\providecommand \natexlab [1]{#1}%
\providecommand \enquote  [1]{``#1''}%
\providecommand \bibnamefont  [1]{#1}%
\providecommand \bibfnamefont [1]{#1}%
\providecommand \citenamefont [1]{#1}%
\providecommand \href@noop [0]{\@secondoftwo}%
\providecommand \href [0]{\begingroup \@sanitize@url \@href}%
\providecommand \@href[1]{\@@startlink{#1}\@@href}%
\providecommand \@@href[1]{\endgroup#1\@@endlink}%
\providecommand \@sanitize@url [0]{\catcode `\\12\catcode `\$12\catcode `\&12\catcode `\#12\catcode `\^12\catcode `\_12\catcode `\%12\relax}%
\providecommand \@@startlink[1]{}%
\providecommand \@@endlink[0]{}%
\providecommand \url  [0]{\begingroup\@sanitize@url \@url }%
\providecommand \@url [1]{\endgroup\@href {#1}{\urlprefix }}%
\providecommand \urlprefix  [0]{URL }%
\providecommand \Eprint [0]{\href }%
\providecommand \doibase [0]{https://doi.org/}%
\providecommand \selectlanguage [0]{\@gobble}%
\providecommand \bibinfo  [0]{\@secondoftwo}%
\providecommand \bibfield  [0]{\@secondoftwo}%
\providecommand \translation [1]{[#1]}%
\providecommand \BibitemOpen [0]{}%
\providecommand \bibitemStop [0]{}%
\providecommand \bibitemNoStop [0]{.\EOS\space}%
\providecommand \EOS [0]{\spacefactor3000\relax}%
\providecommand \BibitemShut  [1]{\csname bibitem#1\endcsname}%
\let\auto@bib@innerbib\@empty
\bibitem [{\citenamefont {Simmonds}\ \emph {et~al.}(2004)\citenamefont {Simmonds}, \citenamefont {Lang}, \citenamefont {Hite}, \citenamefont {Nam}, \citenamefont {Pappas},\ and\ \citenamefont {Martinis}}]{Simmonds.Phys.Rev.Lett.93.077003}%
  \BibitemOpen
  \bibfield  {author} {\bibinfo {author} {\bibfnamefont {R.~W.}\ \bibnamefont {Simmonds}}, \bibinfo {author} {\bibfnamefont {K.~M.}\ \bibnamefont {Lang}}, \bibinfo {author} {\bibfnamefont {D.~A.}\ \bibnamefont {Hite}}, \bibinfo {author} {\bibfnamefont {S.}~\bibnamefont {Nam}}, \bibinfo {author} {\bibfnamefont {D.~P.}\ \bibnamefont {Pappas}},\ and\ \bibinfo {author} {\bibfnamefont {J.~M.}\ \bibnamefont {Martinis}},\ }\bibfield  {title} {\bibinfo {title} {Decoherence in josephson phase qubits from junction resonators},\ }\href {https://doi.org/10.1103/PhysRevLett.93.077003} {\bibfield  {journal} {\bibinfo  {journal} {Phys. Rev. Lett.}\ }\textbf {\bibinfo {volume} {93}},\ \bibinfo {pages} {077003} (\bibinfo {year} {2004})}\BibitemShut {NoStop}%
\bibitem [{\citenamefont {Martinis}\ \emph {et~al.}(2005)\citenamefont {Martinis}, \citenamefont {Cooper}, \citenamefont {McDermott}, \citenamefont {Steffen}, \citenamefont {Ansmann}, \citenamefont {Osborn}, \citenamefont {Cicak}, \citenamefont {Oh}, \citenamefont {Pappas}, \citenamefont {Simmonds},\ and\ \citenamefont {Yu}}]{Martinis.Phys.Rev.Lett.95.210503}%
  \BibitemOpen
  \bibfield  {author} {\bibinfo {author} {\bibfnamefont {J.~M.}\ \bibnamefont {Martinis}}, \bibinfo {author} {\bibfnamefont {K.~B.}\ \bibnamefont {Cooper}}, \bibinfo {author} {\bibfnamefont {R.}~\bibnamefont {McDermott}}, \bibinfo {author} {\bibfnamefont {M.}~\bibnamefont {Steffen}}, \bibinfo {author} {\bibfnamefont {M.}~\bibnamefont {Ansmann}}, \bibinfo {author} {\bibfnamefont {K.~D.}\ \bibnamefont {Osborn}}, \bibinfo {author} {\bibfnamefont {K.}~\bibnamefont {Cicak}}, \bibinfo {author} {\bibfnamefont {S.}~\bibnamefont {Oh}}, \bibinfo {author} {\bibfnamefont {D.~P.}\ \bibnamefont {Pappas}}, \bibinfo {author} {\bibfnamefont {R.~W.}\ \bibnamefont {Simmonds}},\ and\ \bibinfo {author} {\bibfnamefont {C.~C.}\ \bibnamefont {Yu}},\ }\bibfield  {title} {\bibinfo {title} {Decoherence in josephson qubits from dielectric loss},\ }\href {https://doi.org/10.1103/PhysRevLett.95.210503} {\bibfield  {journal} {\bibinfo  {journal} {Phys. Rev. Lett.}\ }\textbf {\bibinfo {volume} {95}},\ \bibinfo {pages} {210503} (\bibinfo {year} {2005})}\BibitemShut {NoStop}%
\bibitem [{\citenamefont {Weeden}\ \emph {et~al.}(2025)\citenamefont {Weeden}, \citenamefont {Harrison}, \citenamefont {Patel}, \citenamefont {Snyder}, \citenamefont {Blackwell}, \citenamefont {Spahn}, \citenamefont {Abdullah}, \citenamefont {Takeda}, \citenamefont {Plourde}, \citenamefont {Martinis},\ and\ \citenamefont {McDermott}}]{Weeden.ax.2506.00193}%
  \BibitemOpen
  \bibfield  {author} {\bibinfo {author} {\bibfnamefont {S.}~\bibnamefont {Weeden}}, \bibinfo {author} {\bibfnamefont {D.~C.}\ \bibnamefont {Harrison}}, \bibinfo {author} {\bibfnamefont {S.}~\bibnamefont {Patel}}, \bibinfo {author} {\bibfnamefont {M.}~\bibnamefont {Snyder}}, \bibinfo {author} {\bibfnamefont {E.~J.}\ \bibnamefont {Blackwell}}, \bibinfo {author} {\bibfnamefont {G.}~\bibnamefont {Spahn}}, \bibinfo {author} {\bibfnamefont {S.}~\bibnamefont {Abdullah}}, \bibinfo {author} {\bibfnamefont {Y.}~\bibnamefont {Takeda}}, \bibinfo {author} {\bibfnamefont {B.~L.~T.}\ \bibnamefont {Plourde}}, \bibinfo {author} {\bibfnamefont {J.~M.}\ \bibnamefont {Martinis}},\ and\ \bibinfo {author} {\bibfnamefont {R.}~\bibnamefont {McDermott}},\ }\href {https://arxiv.org/abs/2506.00193} {\bibinfo {title} {Statistics of strongly coupled defects in superconducting qubits}} (\bibinfo {year} {2025}),\ \Eprint {https://arxiv.org/abs/2506.00193} {arXiv:2506.00193 [quant-ph]} \BibitemShut {NoStop}%
\bibitem [{\citenamefont {Grabovskij}\ \emph {et~al.}(2012)\citenamefont {Grabovskij}, \citenamefont {Peichl}, \citenamefont {Lisenfeld}, \citenamefont {Weiss},\ and\ \citenamefont {Ustinov}}]{Grabovskij.Science338.232234}%
  \BibitemOpen
  \bibfield  {author} {\bibinfo {author} {\bibfnamefont {G.~J.}\ \bibnamefont {Grabovskij}}, \bibinfo {author} {\bibfnamefont {T.}~\bibnamefont {Peichl}}, \bibinfo {author} {\bibfnamefont {J.}~\bibnamefont {Lisenfeld}}, \bibinfo {author} {\bibfnamefont {G.}~\bibnamefont {Weiss}},\ and\ \bibinfo {author} {\bibfnamefont {A.~V.}\ \bibnamefont {Ustinov}},\ }\bibfield  {title} {\bibinfo {title} {Strain tuning of individual atomic tunneling systems detected by a superconducting qubit},\ }\href {https://doi.org/10.1126/science.1226487} {\bibfield  {journal} {\bibinfo  {journal} {Science}\ }\textbf {\bibinfo {volume} {338}},\ \bibinfo {pages} {232} (\bibinfo {year} {2012})}\BibitemShut {NoStop}%
\bibitem [{\citenamefont {Lisenfeld}\ \emph {et~al.}(2019)\citenamefont {Lisenfeld}, \citenamefont {Bilmes}, \citenamefont {Megrant}, \citenamefont {Barends}, \citenamefont {Kelly}, \citenamefont {Klimov}, \citenamefont {Weiss}, \citenamefont {Martinis},\ and\ \citenamefont {Ustinov}}]{Lisenfeld.npjQuantumInformation5.105}%
  \BibitemOpen
  \bibfield  {author} {\bibinfo {author} {\bibfnamefont {J.}~\bibnamefont {Lisenfeld}}, \bibinfo {author} {\bibfnamefont {A.}~\bibnamefont {Bilmes}}, \bibinfo {author} {\bibfnamefont {A.}~\bibnamefont {Megrant}}, \bibinfo {author} {\bibfnamefont {R.}~\bibnamefont {Barends}}, \bibinfo {author} {\bibfnamefont {J.}~\bibnamefont {Kelly}}, \bibinfo {author} {\bibfnamefont {P.}~\bibnamefont {Klimov}}, \bibinfo {author} {\bibfnamefont {G.}~\bibnamefont {Weiss}}, \bibinfo {author} {\bibfnamefont {J.~M.}\ \bibnamefont {Martinis}},\ and\ \bibinfo {author} {\bibfnamefont {A.~V.}\ \bibnamefont {Ustinov}},\ }\bibfield  {title} {\bibinfo {title} {Electric field spectroscopy of material defects in transmon qubits},\ }\href {https://doi.org/10.1038/s41534-019-0224-1} {\bibfield  {journal} {\bibinfo  {journal} {npj Quantum Information}\ }\textbf {\bibinfo {volume} {5}},\ \bibinfo {pages} {105} (\bibinfo {year} {2019})}\BibitemShut {NoStop}%
\bibitem [{\citenamefont {Kline}\ \emph {et~al.}(2008)\citenamefont {Kline}, \citenamefont {Wang}, \citenamefont {Oh}, \citenamefont {Martinis},\ and\ \citenamefont {Pappas}}]{Kline.SST.22.015004}%
  \BibitemOpen
  \bibfield  {author} {\bibinfo {author} {\bibfnamefont {J.~S.}\ \bibnamefont {Kline}}, \bibinfo {author} {\bibfnamefont {H.}~\bibnamefont {Wang}}, \bibinfo {author} {\bibfnamefont {S.}~\bibnamefont {Oh}}, \bibinfo {author} {\bibfnamefont {J.~M.}\ \bibnamefont {Martinis}},\ and\ \bibinfo {author} {\bibfnamefont {D.~P.}\ \bibnamefont {Pappas}},\ }\bibfield  {title} {\bibinfo {title} {Josephson phase qubit circuit for the evaluation of advanced tunnel barrier materials},\ }\href {https://doi.org/10.1088/0953-2048/22/1/015004} {\bibfield  {journal} {\bibinfo  {journal} {Superconductor Science and Technology}\ }\textbf {\bibinfo {volume} {22}},\ \bibinfo {pages} {015004} (\bibinfo {year} {2008})}\BibitemShut {NoStop}%
\bibitem [{\citenamefont {Oh}\ \emph {et~al.}(2006)\citenamefont {Oh}, \citenamefont {Cicak}, \citenamefont {Kline}, \citenamefont {Sillanp\"a\"a}, \citenamefont {Osborn}, \citenamefont {Whittaker}, \citenamefont {Simmonds},\ and\ \citenamefont {Pappas}}]{Oh.PRB.74.100502}%
  \BibitemOpen
  \bibfield  {author} {\bibinfo {author} {\bibfnamefont {S.}~\bibnamefont {Oh}}, \bibinfo {author} {\bibfnamefont {K.}~\bibnamefont {Cicak}}, \bibinfo {author} {\bibfnamefont {J.~S.}\ \bibnamefont {Kline}}, \bibinfo {author} {\bibfnamefont {M.~A.}\ \bibnamefont {Sillanp\"a\"a}}, \bibinfo {author} {\bibfnamefont {K.~D.}\ \bibnamefont {Osborn}}, \bibinfo {author} {\bibfnamefont {J.~D.}\ \bibnamefont {Whittaker}}, \bibinfo {author} {\bibfnamefont {R.~W.}\ \bibnamefont {Simmonds}},\ and\ \bibinfo {author} {\bibfnamefont {D.~P.}\ \bibnamefont {Pappas}},\ }\bibfield  {title} {\bibinfo {title} {Elimination of two level fluctuators in superconducting quantum bits by an epitaxial tunnel barrier},\ }\href {https://doi.org/10.1103/PhysRevB.74.100502} {\bibfield  {journal} {\bibinfo  {journal} {Phys. Rev. B}\ }\textbf {\bibinfo {volume} {74}},\ \bibinfo {pages} {100502} (\bibinfo {year} {2006})}\BibitemShut {NoStop}%
\bibitem [{\citenamefont {Anferov}\ \emph {et~al.}(2024)\citenamefont {Anferov}, \citenamefont {Lee}, \citenamefont {Zhao}, \citenamefont {Simon},\ and\ \citenamefont {Schuster}}]{Anferov.PRApp.21.024047}%
  \BibitemOpen
  \bibfield  {author} {\bibinfo {author} {\bibfnamefont {A.}~\bibnamefont {Anferov}}, \bibinfo {author} {\bibfnamefont {K.-H.}\ \bibnamefont {Lee}}, \bibinfo {author} {\bibfnamefont {F.}~\bibnamefont {Zhao}}, \bibinfo {author} {\bibfnamefont {J.}~\bibnamefont {Simon}},\ and\ \bibinfo {author} {\bibfnamefont {D.~I.}\ \bibnamefont {Schuster}},\ }\bibfield  {title} {\bibinfo {title} {Improved coherence in optically defined niobium trilayer-junction qubits},\ }\href {https://doi.org/10.1103/PhysRevApplied.21.024047} {\bibfield  {journal} {\bibinfo  {journal} {Phys. Rev. Appl.}\ }\textbf {\bibinfo {volume} {21}},\ \bibinfo {pages} {024047} (\bibinfo {year} {2024})}\BibitemShut {NoStop}%
\bibitem [{\citenamefont {Kim}\ \emph {et~al.}(2021)\citenamefont {Kim}, \citenamefont {Terai}, \citenamefont {Yamashita}, \citenamefont {Qiu}, \citenamefont {Fuse}, \citenamefont {Yoshihara}, \citenamefont {Ashhab}, \citenamefont {Inomata},\ and\ \citenamefont {Semba}}]{Kim.CommMat.2.98}%
  \BibitemOpen
  \bibfield  {author} {\bibinfo {author} {\bibfnamefont {S.}~\bibnamefont {Kim}}, \bibinfo {author} {\bibfnamefont {H.}~\bibnamefont {Terai}}, \bibinfo {author} {\bibfnamefont {T.}~\bibnamefont {Yamashita}}, \bibinfo {author} {\bibfnamefont {W.}~\bibnamefont {Qiu}}, \bibinfo {author} {\bibfnamefont {T.}~\bibnamefont {Fuse}}, \bibinfo {author} {\bibfnamefont {F.}~\bibnamefont {Yoshihara}}, \bibinfo {author} {\bibfnamefont {S.}~\bibnamefont {Ashhab}}, \bibinfo {author} {\bibfnamefont {K.}~\bibnamefont {Inomata}},\ and\ \bibinfo {author} {\bibfnamefont {K.}~\bibnamefont {Semba}},\ }\bibfield  {title} {\bibinfo {title} {Enhanced coherence of all-nitride superconducting qubits epitaxially grown on silicon substrate},\ }\href {https://doi.org/10.1038/s43246-021-00204-4} {\bibfield  {journal} {\bibinfo  {journal} {Communications Materials}\ }\textbf {\bibinfo {volume} {2}},\ \bibinfo {pages} {98} (\bibinfo {year} {2021})}\BibitemShut {NoStop}%
\bibitem [{\citenamefont {Van~Damme}\ \emph {et~al.}(2024)\citenamefont {Van~Damme}, \citenamefont {Massar}, \citenamefont {Acharya}, \citenamefont {Ivanov}, \citenamefont {Perez~Lozano}, \citenamefont {Canvel}, \citenamefont {Demarets}, \citenamefont {Vangoidsenhoven}, \citenamefont {Hermans}, \citenamefont {Lai}, \citenamefont {Vadiraj}, \citenamefont {Mongillo}, \citenamefont {Wan}, \citenamefont {De~Boeck}, \citenamefont {Poto{\v{c}}nik},\ and\ \citenamefont {De~Greve}}]{VanDamme.Nature.634.7479}%
  \BibitemOpen
  \bibfield  {author} {\bibinfo {author} {\bibfnamefont {J.}~\bibnamefont {Van~Damme}}, \bibinfo {author} {\bibfnamefont {S.}~\bibnamefont {Massar}}, \bibinfo {author} {\bibfnamefont {R.}~\bibnamefont {Acharya}}, \bibinfo {author} {\bibfnamefont {T.}~\bibnamefont {Ivanov}}, \bibinfo {author} {\bibfnamefont {D.}~\bibnamefont {Perez~Lozano}}, \bibinfo {author} {\bibfnamefont {Y.}~\bibnamefont {Canvel}}, \bibinfo {author} {\bibfnamefont {M.}~\bibnamefont {Demarets}}, \bibinfo {author} {\bibfnamefont {D.}~\bibnamefont {Vangoidsenhoven}}, \bibinfo {author} {\bibfnamefont {Y.}~\bibnamefont {Hermans}}, \bibinfo {author} {\bibfnamefont {J.~G.}\ \bibnamefont {Lai}}, \bibinfo {author} {\bibfnamefont {A.~M.}\ \bibnamefont {Vadiraj}}, \bibinfo {author} {\bibfnamefont {M.}~\bibnamefont {Mongillo}}, \bibinfo {author} {\bibfnamefont {D.}~\bibnamefont {Wan}}, \bibinfo {author} {\bibfnamefont {J.}~\bibnamefont {De~Boeck}}, \bibinfo {author} {\bibfnamefont {A.}~\bibnamefont {Poto{\v{c}}nik}},\ and\ \bibinfo {author} {\bibfnamefont {K.}~\bibnamefont {De~Greve}},\ }\bibfield  {title} {\bibinfo {title} {Advanced cmos manufacturing of superconducting qubits on 300 mm wafers},\ }\href {https://doi.org/10.1038/s41586-024-07941-9} {\bibfield  {journal} {\bibinfo  {journal} {Nature}\ }\textbf {\bibinfo {volume} {634}},\ \bibinfo {pages} {74} (\bibinfo {year} {2024})}\BibitemShut {NoStop}%
\bibitem [{\citenamefont {{Potts}}\ \emph {et~al.}(2001)\citenamefont {{Potts}}, \citenamefont {{Parker}}, \citenamefont {{Baumberg}},\ and\ \citenamefont {{de Groot}}}]{Potts.IEEProceedingsScienceMeasurementandTechnology148.225228}%
  \BibitemOpen
  \bibfield  {author} {\bibinfo {author} {\bibfnamefont {A.}~\bibnamefont {{Potts}}}, \bibinfo {author} {\bibfnamefont {G.~J.}\ \bibnamefont {{Parker}}}, \bibinfo {author} {\bibfnamefont {J.~J.}\ \bibnamefont {{Baumberg}}},\ and\ \bibinfo {author} {\bibfnamefont {P.~A.~J.}\ \bibnamefont {{de Groot}}},\ }\bibfield  {title} {\bibinfo {title} {Cmos compatible fabrication methods for submicron josephson junction qubits},\ }\href {https://doi.org/10.1049/ip-smt:20010395} {\bibfield  {journal} {\bibinfo  {journal} {IEE Proceedings - Science, Measurement and Technology}\ }\textbf {\bibinfo {volume} {148}},\ \bibinfo {pages} {225} (\bibinfo {year} {2001})}\BibitemShut {NoStop}%
\bibitem [{\citenamefont {Megrant}\ \emph {et~al.}(2012)\citenamefont {Megrant}, \citenamefont {Neill}, \citenamefont {Barends}, \citenamefont {Chiaro}, \citenamefont {Chen}, \citenamefont {Feigl}, \citenamefont {Kelly}, \citenamefont {Lucero}, \citenamefont {Mariantoni}, \citenamefont {O’Malley}, \citenamefont {Sank}, \citenamefont {Vainsencher}, \citenamefont {Wenner}, \citenamefont {White}, \citenamefont {Yin}, \citenamefont {Zhao}, \citenamefont {Palmstrøm}, \citenamefont {Martinis},\ and\ \citenamefont {Cleland}}]{Megrant.AppliedPhysicsLetters100.113510}%
  \BibitemOpen
  \bibfield  {author} {\bibinfo {author} {\bibfnamefont {A.}~\bibnamefont {Megrant}}, \bibinfo {author} {\bibfnamefont {C.}~\bibnamefont {Neill}}, \bibinfo {author} {\bibfnamefont {R.}~\bibnamefont {Barends}}, \bibinfo {author} {\bibfnamefont {B.}~\bibnamefont {Chiaro}}, \bibinfo {author} {\bibfnamefont {Y.}~\bibnamefont {Chen}}, \bibinfo {author} {\bibfnamefont {L.}~\bibnamefont {Feigl}}, \bibinfo {author} {\bibfnamefont {J.}~\bibnamefont {Kelly}}, \bibinfo {author} {\bibfnamefont {E.}~\bibnamefont {Lucero}}, \bibinfo {author} {\bibfnamefont {M.}~\bibnamefont {Mariantoni}}, \bibinfo {author} {\bibfnamefont {P.~J.~J.}\ \bibnamefont {O’Malley}}, \bibinfo {author} {\bibfnamefont {D.}~\bibnamefont {Sank}}, \bibinfo {author} {\bibfnamefont {A.}~\bibnamefont {Vainsencher}}, \bibinfo {author} {\bibfnamefont {J.}~\bibnamefont {Wenner}}, \bibinfo {author} {\bibfnamefont {T.~C.}\ \bibnamefont {White}}, \bibinfo {author} {\bibfnamefont {Y.}~\bibnamefont {Yin}}, \bibinfo {author} {\bibfnamefont {J.}~\bibnamefont {Zhao}}, \bibinfo {author} {\bibfnamefont {C.~J.}\ \bibnamefont {Palmstrøm}}, \bibinfo {author} {\bibfnamefont {J.~M.}\ \bibnamefont {Martinis}},\ and\ \bibinfo {author} {\bibfnamefont {A.~N.}\ \bibnamefont {Cleland}},\ }\bibfield  {title} {\bibinfo {title} {Planar superconducting resonators with internal quality factors above one million},\ }\href {https://doi.org/10.1063/1.3693409} {\bibfield  {journal} {\bibinfo  {journal} {Applied Physics Letters}\ }\textbf {\bibinfo {volume} {100}},\ \bibinfo {pages} {113510} (\bibinfo {year} {2012})}\BibitemShut {NoStop}%
\bibitem [{\citenamefont {McRae}\ \emph {et~al.}(2020)\citenamefont {McRae}, \citenamefont {Wang}, \citenamefont {Gao}, \citenamefont {Vissers}, \citenamefont {Brecht}, \citenamefont {Dunsworth}, \citenamefont {Pappas},\ and\ \citenamefont {Mutus}}]{McRae.ReviewofScientificInstruments91.091101}%
  \BibitemOpen
  \bibfield  {author} {\bibinfo {author} {\bibfnamefont {C.~R.~H.}\ \bibnamefont {McRae}}, \bibinfo {author} {\bibfnamefont {H.}~\bibnamefont {Wang}}, \bibinfo {author} {\bibfnamefont {J.}~\bibnamefont {Gao}}, \bibinfo {author} {\bibfnamefont {M.~R.}\ \bibnamefont {Vissers}}, \bibinfo {author} {\bibfnamefont {T.}~\bibnamefont {Brecht}}, \bibinfo {author} {\bibfnamefont {A.}~\bibnamefont {Dunsworth}}, \bibinfo {author} {\bibfnamefont {D.~P.}\ \bibnamefont {Pappas}},\ and\ \bibinfo {author} {\bibfnamefont {J.}~\bibnamefont {Mutus}},\ }\bibfield  {title} {\bibinfo {title} {Materials loss measurements using superconducting microwave resonators},\ }\href {https://doi.org/10.1063/5.0017378} {\bibfield  {journal} {\bibinfo  {journal} {Review of Scientific Instruments}\ }\textbf {\bibinfo {volume} {91}},\ \bibinfo {pages} {091101} (\bibinfo {year} {2020})},\ \Eprint {https://arxiv.org/abs/https://doi.org/10.1063/5.0017378} {https://doi.org/10.1063/5.0017378} \BibitemShut {NoStop}%
\bibitem [{Sup()}]{Supp.Mate.}%
  \BibitemOpen
  \href@noop {} {}\bibinfo {note} {Supplementary Material is available at this link.}\BibitemShut {Stop}%
\bibitem [{\citenamefont {Burnett}\ \emph {et~al.}(2018)\citenamefont {Burnett}, \citenamefont {Bengtsson}, \citenamefont {Niepce},\ and\ \citenamefont {Bylander}}]{Burnett.JoPCS.969.012131}%
  \BibitemOpen
  \bibfield  {author} {\bibinfo {author} {\bibfnamefont {J.}~\bibnamefont {Burnett}}, \bibinfo {author} {\bibfnamefont {A.}~\bibnamefont {Bengtsson}}, \bibinfo {author} {\bibfnamefont {D.}~\bibnamefont {Niepce}},\ and\ \bibinfo {author} {\bibfnamefont {J.}~\bibnamefont {Bylander}},\ }\bibfield  {title} {\bibinfo {title} {Noise and loss of superconducting aluminium resonators at single photon energies},\ }\href {https://doi.org/10.1088/1742-6596/969/1/012131} {\bibfield  {journal} {\bibinfo  {journal} {Journal of Physics: Conference Series}\ }\textbf {\bibinfo {volume} {969}},\ \bibinfo {pages} {012131} (\bibinfo {year} {2018})}\BibitemShut {NoStop}%
\bibitem [{\citenamefont {Gao}(2008)}]{Gao.Thesis.2008}%
  \BibitemOpen
  \bibfield  {author} {\bibinfo {author} {\bibfnamefont {J.}~\bibnamefont {Gao}},\ }\emph {\bibinfo {title} {The Physics of Superconducting Microwave Resonators}},\ \href {https://uwspace.uwaterloo.ca/handle/10012/9575} {Ph.D. thesis},\ \bibinfo  {school} {California Institute of Technology} (\bibinfo {year} {2008})\BibitemShut {NoStop}%
\bibitem [{\citenamefont {Wang}\ \emph {et~al.}(2015)\citenamefont {Wang}, \citenamefont {Axline}, \citenamefont {Gao}, \citenamefont {Brecht}, \citenamefont {Chu}, \citenamefont {Frunzio}, \citenamefont {Devoret},\ and\ \citenamefont {Schoelkopf}}]{Wang.APL.107.162601}%
  \BibitemOpen
  \bibfield  {author} {\bibinfo {author} {\bibfnamefont {C.}~\bibnamefont {Wang}}, \bibinfo {author} {\bibfnamefont {C.}~\bibnamefont {Axline}}, \bibinfo {author} {\bibfnamefont {Y.~Y.}\ \bibnamefont {Gao}}, \bibinfo {author} {\bibfnamefont {T.}~\bibnamefont {Brecht}}, \bibinfo {author} {\bibfnamefont {Y.}~\bibnamefont {Chu}}, \bibinfo {author} {\bibfnamefont {L.}~\bibnamefont {Frunzio}}, \bibinfo {author} {\bibfnamefont {M.~H.}\ \bibnamefont {Devoret}},\ and\ \bibinfo {author} {\bibfnamefont {R.~J.}\ \bibnamefont {Schoelkopf}},\ }\bibfield  {title} {\bibinfo {title} {Surface participation and dielectric loss in superconducting qubits},\ }\href {https://doi.org/10.1063/1.4934486} {\bibfield  {journal} {\bibinfo  {journal} {Applied Physics Letters}\ }\textbf {\bibinfo {volume} {107}},\ \bibinfo {pages} {162601} (\bibinfo {year} {2015})},\ \Eprint {https://arxiv.org/abs/https://doi.org/10.1063/1.4934486} {https://doi.org/10.1063/1.4934486} \BibitemShut {NoStop}%
\bibitem [{\citenamefont {Bilmes}\ \emph {et~al.}(2022)\citenamefont {Bilmes}, \citenamefont {Volosheniuk}, \citenamefont {Ustinov},\ and\ \citenamefont {Lisenfeld}}]{Bilmes.NPJQI.8.24}%
  \BibitemOpen
  \bibfield  {author} {\bibinfo {author} {\bibfnamefont {A.}~\bibnamefont {Bilmes}}, \bibinfo {author} {\bibfnamefont {S.}~\bibnamefont {Volosheniuk}}, \bibinfo {author} {\bibfnamefont {A.~V.}\ \bibnamefont {Ustinov}},\ and\ \bibinfo {author} {\bibfnamefont {J.}~\bibnamefont {Lisenfeld}},\ }\bibfield  {title} {\bibinfo {title} {Probing defect densities at the edges and inside josephson junctions of superconducting qubits},\ }\href {https://doi.org/10.1038/s41534-022-00532-4} {\bibfield  {journal} {\bibinfo  {journal} {npj Quantum Information}\ }\textbf {\bibinfo {volume} {8}},\ \bibinfo {pages} {24} (\bibinfo {year} {2022})}\BibitemShut {NoStop}%
\bibitem [{\citenamefont {Colao~Zanuz}\ \emph {et~al.}(2025)\citenamefont {Colao~Zanuz}, \citenamefont {Ficheux}, \citenamefont {Michaud}, \citenamefont {Orekhov}, \citenamefont {Hanke}, \citenamefont {Flasby}, \citenamefont {Bahrami~Panah}, \citenamefont {Norris}, \citenamefont {Kerschbaum}, \citenamefont {Remm}, \citenamefont {Swiadek}, \citenamefont {Hellings}, \citenamefont {Laz\ifmmode~\u{a}\else \u{a}\fi{}r}, \citenamefont {Scarato}, \citenamefont {Lacroix}, \citenamefont {Krinner}, \citenamefont {Eichler}, \citenamefont {Wallraff},\ and\ \citenamefont {Besse}}]{ColaoZanuz.PRApp.23.044054}%
  \BibitemOpen
  \bibfield  {author} {\bibinfo {author} {\bibfnamefont {D.}~\bibnamefont {Colao~Zanuz}}, \bibinfo {author} {\bibfnamefont {Q.}~\bibnamefont {Ficheux}}, \bibinfo {author} {\bibfnamefont {L.}~\bibnamefont {Michaud}}, \bibinfo {author} {\bibfnamefont {A.}~\bibnamefont {Orekhov}}, \bibinfo {author} {\bibfnamefont {K.}~\bibnamefont {Hanke}}, \bibinfo {author} {\bibfnamefont {A.}~\bibnamefont {Flasby}}, \bibinfo {author} {\bibfnamefont {M.}~\bibnamefont {Bahrami~Panah}}, \bibinfo {author} {\bibfnamefont {G.~J.}\ \bibnamefont {Norris}}, \bibinfo {author} {\bibfnamefont {M.}~\bibnamefont {Kerschbaum}}, \bibinfo {author} {\bibfnamefont {A.}~\bibnamefont {Remm}}, \bibinfo {author} {\bibfnamefont {F.~m.~c.}\ \bibnamefont {Swiadek}}, \bibinfo {author} {\bibfnamefont {C.}~\bibnamefont {Hellings}}, \bibinfo {author} {\bibfnamefont {S.}~\bibnamefont {Laz\ifmmode~\u{a}\else \u{a}\fi{}r}}, \bibinfo {author} {\bibfnamefont {C.}~\bibnamefont {Scarato}}, \bibinfo {author} {\bibfnamefont {N.}~\bibnamefont {Lacroix}}, \bibinfo {author} {\bibfnamefont {S.}~\bibnamefont {Krinner}}, \bibinfo {author} {\bibfnamefont {C.}~\bibnamefont {Eichler}}, \bibinfo {author} {\bibfnamefont {A.}~\bibnamefont {Wallraff}},\ and\ \bibinfo {author} {\bibfnamefont {J.-C.}\ \bibnamefont {Besse}},\ }\bibfield  {title} {\bibinfo {title} {Mitigating losses of superconducting qubits strongly coupled to defect modes},\ }\href {https://doi.org/10.1103/PhysRevApplied.23.044054} {\bibfield  {journal} {\bibinfo  {journal} {Phys. Rev. Appl.}\ }\textbf {\bibinfo {volume} {23}},\ \bibinfo {pages} {044054} (\bibinfo {year} {2025})}\BibitemShut {NoStop}%
\bibitem [{\citenamefont {Abdurakhimov}\ \emph {et~al.}(2022)\citenamefont {Abdurakhimov}, \citenamefont {Mahboob}, \citenamefont {Toida}, \citenamefont {Kakuyanagi}, \citenamefont {Matsuzaki},\ and\ \citenamefont {Saito}}]{Abdurakhimov.PRXQ.3.040332}%
  \BibitemOpen
  \bibfield  {author} {\bibinfo {author} {\bibfnamefont {L.~V.}\ \bibnamefont {Abdurakhimov}}, \bibinfo {author} {\bibfnamefont {I.}~\bibnamefont {Mahboob}}, \bibinfo {author} {\bibfnamefont {H.}~\bibnamefont {Toida}}, \bibinfo {author} {\bibfnamefont {K.}~\bibnamefont {Kakuyanagi}}, \bibinfo {author} {\bibfnamefont {Y.}~\bibnamefont {Matsuzaki}},\ and\ \bibinfo {author} {\bibfnamefont {S.}~\bibnamefont {Saito}},\ }\bibfield  {title} {\bibinfo {title} {Identification of different types of high-frequency defects in superconducting qubits},\ }\href {https://doi.org/10.1103/PRXQuantum.3.040332} {\bibfield  {journal} {\bibinfo  {journal} {PRX Quantum}\ }\textbf {\bibinfo {volume} {3}},\ \bibinfo {pages} {040332} (\bibinfo {year} {2022})}\BibitemShut {NoStop}%
\bibitem [{\citenamefont {Braum\"uller}\ \emph {et~al.}(2020)\citenamefont {Braum\"uller}, \citenamefont {Ding}, \citenamefont {Veps\"al\"ainen}, \citenamefont {Sung}, \citenamefont {Kjaergaard}, \citenamefont {Menke}, \citenamefont {Winik}, \citenamefont {Kim}, \citenamefont {Niedzielski}, \citenamefont {Melville}, \citenamefont {Yoder}, \citenamefont {Hirjibehedin}, \citenamefont {Orlando}, \citenamefont {Gustavsson},\ and\ \citenamefont {Oliver}}]{Braumueller.PRA.13.054079}%
  \BibitemOpen
  \bibfield  {author} {\bibinfo {author} {\bibfnamefont {J.}~\bibnamefont {Braum\"uller}}, \bibinfo {author} {\bibfnamefont {L.}~\bibnamefont {Ding}}, \bibinfo {author} {\bibfnamefont {A.~P.}\ \bibnamefont {Veps\"al\"ainen}}, \bibinfo {author} {\bibfnamefont {Y.}~\bibnamefont {Sung}}, \bibinfo {author} {\bibfnamefont {M.}~\bibnamefont {Kjaergaard}}, \bibinfo {author} {\bibfnamefont {T.}~\bibnamefont {Menke}}, \bibinfo {author} {\bibfnamefont {R.}~\bibnamefont {Winik}}, \bibinfo {author} {\bibfnamefont {D.}~\bibnamefont {Kim}}, \bibinfo {author} {\bibfnamefont {B.~M.}\ \bibnamefont {Niedzielski}}, \bibinfo {author} {\bibfnamefont {A.}~\bibnamefont {Melville}}, \bibinfo {author} {\bibfnamefont {J.~L.}\ \bibnamefont {Yoder}}, \bibinfo {author} {\bibfnamefont {C.~F.}\ \bibnamefont {Hirjibehedin}}, \bibinfo {author} {\bibfnamefont {T.~P.}\ \bibnamefont {Orlando}}, \bibinfo {author} {\bibfnamefont {S.}~\bibnamefont {Gustavsson}},\ and\ \bibinfo {author} {\bibfnamefont {W.~D.}\ \bibnamefont {Oliver}},\ }\bibfield  {title} {\bibinfo {title} {Characterizing and optimizing qubit coherence based on squid geometry},\ }\href {https://doi.org/10.1103/PhysRevApplied.13.054079} {\bibfield  {journal} {\bibinfo  {journal} {Phys. Rev. Applied}\ }\textbf {\bibinfo {volume} {13}},\ \bibinfo {pages} {054079} (\bibinfo {year} {2020})}\BibitemShut {NoStop}%
\bibitem [{\citenamefont {Deng}\ \emph {et~al.}(2023)\citenamefont {Deng}, \citenamefont {Song}, \citenamefont {Gao}, \citenamefont {Xia}, \citenamefont {Bao}, \citenamefont {Jiang}, \citenamefont {Ku}, \citenamefont {Li}, \citenamefont {Ma}, \citenamefont {Qin}, \citenamefont {Sun}, \citenamefont {Tang}, \citenamefont {Wang}, \citenamefont {Wu}, \citenamefont {Yu}, \citenamefont {Zhang}, \citenamefont {Zhang}, \citenamefont {Zhou}, \citenamefont {Zhu}, \citenamefont {Shi}, \citenamefont {Zhao},\ and\ \citenamefont {Deng}}]{Deng.PRA.19.024013}%
  \BibitemOpen
  \bibfield  {author} {\bibinfo {author} {\bibfnamefont {H.}~\bibnamefont {Deng}}, \bibinfo {author} {\bibfnamefont {Z.}~\bibnamefont {Song}}, \bibinfo {author} {\bibfnamefont {R.}~\bibnamefont {Gao}}, \bibinfo {author} {\bibfnamefont {T.}~\bibnamefont {Xia}}, \bibinfo {author} {\bibfnamefont {F.}~\bibnamefont {Bao}}, \bibinfo {author} {\bibfnamefont {X.}~\bibnamefont {Jiang}}, \bibinfo {author} {\bibfnamefont {H.-S.}\ \bibnamefont {Ku}}, \bibinfo {author} {\bibfnamefont {Z.}~\bibnamefont {Li}}, \bibinfo {author} {\bibfnamefont {X.}~\bibnamefont {Ma}}, \bibinfo {author} {\bibfnamefont {J.}~\bibnamefont {Qin}}, \bibinfo {author} {\bibfnamefont {H.}~\bibnamefont {Sun}}, \bibinfo {author} {\bibfnamefont {C.}~\bibnamefont {Tang}}, \bibinfo {author} {\bibfnamefont {T.}~\bibnamefont {Wang}}, \bibinfo {author} {\bibfnamefont {F.}~\bibnamefont {Wu}}, \bibinfo {author} {\bibfnamefont {W.}~\bibnamefont {Yu}}, \bibinfo {author} {\bibfnamefont {G.}~\bibnamefont {Zhang}}, \bibinfo {author} {\bibfnamefont {X.}~\bibnamefont {Zhang}}, \bibinfo {author} {\bibfnamefont {J.}~\bibnamefont {Zhou}}, \bibinfo {author} {\bibfnamefont {X.}~\bibnamefont {Zhu}}, \bibinfo {author} {\bibfnamefont {Y.}~\bibnamefont {Shi}}, \bibinfo {author} {\bibfnamefont {H.-H.}\ \bibnamefont {Zhao}},\ and\ \bibinfo {author} {\bibfnamefont {C.}~\bibnamefont {Deng}},\ }\bibfield  {title} {\bibinfo {title} {Titanium nitride film on sapphire substrate with low dielectric loss for superconducting qubits},\ }\href {https://doi.org/10.1103/PhysRevApplied.19.024013} {\bibfield  {journal} {\bibinfo  {journal} {Phys. Rev. Appl.}\ }\textbf {\bibinfo {volume} {19}},\ \bibinfo {pages} {024013} (\bibinfo {year} {2023})}\BibitemShut {NoStop}%
\bibitem [{\citenamefont {Wang}\ \emph {et~al.}(2025)\citenamefont {Wang}, \citenamefont {Lu}, \citenamefont {Zhan}, \citenamefont {Ma}, \citenamefont {Wu}, \citenamefont {Sun}, \citenamefont {Deng}, \citenamefont {Bai}, \citenamefont {Bao}, \citenamefont {Chang}, \citenamefont {Gao}, \citenamefont {Gao}, \citenamefont {Gong}, \citenamefont {Hu}, \citenamefont {Hu}, \citenamefont {Ji}, \citenamefont {Ma}, \citenamefont {Mao}, \citenamefont {Song}, \citenamefont {Tang}, \citenamefont {Wang}, \citenamefont {Wang}, \citenamefont {Wang}, \citenamefont {Xia}, \citenamefont {Xu}, \citenamefont {Zhan}, \citenamefont {Zhang}, \citenamefont {Zhou}, \citenamefont {Zhu}, \citenamefont {Zhu}, \citenamefont {Zhu}, \citenamefont {Zhu}, \citenamefont {Shi}, \citenamefont {Zhao},\ and\ \citenamefont {Deng}}]{Wang.PRApp.23.044064}%
  \BibitemOpen
  \bibfield  {author} {\bibinfo {author} {\bibfnamefont {F.}~\bibnamefont {Wang}}, \bibinfo {author} {\bibfnamefont {K.}~\bibnamefont {Lu}}, \bibinfo {author} {\bibfnamefont {H.}~\bibnamefont {Zhan}}, \bibinfo {author} {\bibfnamefont {L.}~\bibnamefont {Ma}}, \bibinfo {author} {\bibfnamefont {F.}~\bibnamefont {Wu}}, \bibinfo {author} {\bibfnamefont {H.}~\bibnamefont {Sun}}, \bibinfo {author} {\bibfnamefont {H.}~\bibnamefont {Deng}}, \bibinfo {author} {\bibfnamefont {Y.}~\bibnamefont {Bai}}, \bibinfo {author} {\bibfnamefont {F.}~\bibnamefont {Bao}}, \bibinfo {author} {\bibfnamefont {X.}~\bibnamefont {Chang}}, \bibinfo {author} {\bibfnamefont {R.}~\bibnamefont {Gao}}, \bibinfo {author} {\bibfnamefont {X.}~\bibnamefont {Gao}}, \bibinfo {author} {\bibfnamefont {G.}~\bibnamefont {Gong}}, \bibinfo {author} {\bibfnamefont {L.}~\bibnamefont {Hu}}, \bibinfo {author} {\bibfnamefont {R.}~\bibnamefont {Hu}}, \bibinfo {author} {\bibfnamefont {H.}~\bibnamefont {Ji}}, \bibinfo {author} {\bibfnamefont {X.}~\bibnamefont {Ma}}, \bibinfo {author} {\bibfnamefont {L.}~\bibnamefont {Mao}}, \bibinfo {author} {\bibfnamefont {Z.}~\bibnamefont {Song}}, \bibinfo {author} {\bibfnamefont {C.}~\bibnamefont {Tang}}, \bibinfo {author} {\bibfnamefont {H.}~\bibnamefont {Wang}}, \bibinfo {author} {\bibfnamefont {T.}~\bibnamefont {Wang}}, \bibinfo {author} {\bibfnamefont {Z.}~\bibnamefont {Wang}}, \bibinfo {author} {\bibfnamefont {T.}~\bibnamefont {Xia}}, \bibinfo {author} {\bibfnamefont {H.}~\bibnamefont {Xu}}, \bibinfo {author} {\bibfnamefont {Z.}~\bibnamefont {Zhan}}, \bibinfo {author} {\bibfnamefont {G.}~\bibnamefont {Zhang}}, \bibinfo {author} {\bibfnamefont {T.}~\bibnamefont {Zhou}}, \bibinfo {author} {\bibfnamefont {M.}~\bibnamefont {Zhu}}, \bibinfo {author} {\bibfnamefont {Q.}~\bibnamefont {Zhu}}, \bibinfo {author} {\bibfnamefont {S.}~\bibnamefont {Zhu}}, \bibinfo {author} {\bibfnamefont {X.}~\bibnamefont {Zhu}}, \bibinfo {author} {\bibfnamefont {Y.}~\bibnamefont {Shi}}, \bibinfo {author} {\bibfnamefont {H.-H.}\ \bibnamefont {Zhao}},\ and\ \bibinfo {author} {\bibfnamefont {C.}~\bibnamefont {Deng}},\ }\bibfield  {title} {\bibinfo {title} {High-coherence fluxonium qubits manufactured with a wafer-scale-uniformity process},\ }\href {https://doi.org/10.1103/PhysRevApplied.23.044064} {\bibfield  {journal} {\bibinfo  {journal} {Phys. Rev. Appl.}\ }\textbf {\bibinfo {volume} {23}},\ \bibinfo {pages} {044064} (\bibinfo {year} {2025})}\BibitemShut {NoStop}%
\end{thebibliography}
\end{document}